\documentclass[twocolumn,prb,aps,showpacs,floatfix]{revtex4}

\usepackage{epsfig,url}
\usepackage{epsf}
\usepackage{graphicx}
\usepackage{amsmath}
\usepackage{booktabs}  % nice tables
\usepackage{dcolumn}
\usepackage{rotating}
\usepackage{bm}        % for bold math
\usepackage{bbm}       % for nice three dimensional spaces
\newcommand{\siso}{\delta_{\rm iso}}
\newcommand{\sani}{\delta_{\rm anis}}
\newcommand{\saa}{\delta_{11}}
\newcommand{\sbb}{\delta_{22}}
\newcommand{\scc}{\delta_{33}}

\begin{document}

\title{Magnetic response of carbon nanotubes from {\it ab initio} calculations}

\author{Miguel A.\,L. Marques}
\author{Mayeul d'Avezac}
\author{Francesco Mauri}
 \affiliation{Institut de Min\'eralogie et de Physique des Milieux Condens\'es,
 Universit\'e Paris VI, 140 rue de Lourmel
 75015 Paris, France}

\date{\today}

\begin{abstract}
  We present {\it ab initio} calculations of the magnetic susceptibility and of
  the $^{13}$C chemical shift for carbon nanotubes, both isolated and in
  bundles. These calculations are performed using the recently proposed
  gauge-including projector augmented-wave approach for the calculation of
  magnetic response in periodic insulating systems. We have focused on the
  semiconducting zigzag nanotubes with diameters ranging from 0.6 to 1.6\,nm.
  Both the susceptibility and the isotropic shift exhibit a dependence with the
  diameter (D) and the chirality of the tube (although this dependence is
  stronger for the susceptibility). The isotropic shift behaves asymptotically
  as $\alpha/D + 116.0$, where $\alpha$ is a different constant for each family
  of nanotubes.  For a tube diameter of around 1.2\,nm, a value normally found
  in experimental samples, our results are in excellent agreement with
  experiments. Moreover, we calculated the chemical shift of a double-wall tube.
  We found a diamagnetic shift of the isotropic lines corresponding to the atoms
  of the inner tube due to the effect of the outer tube. This shift is in good
  agreement with recent experiments, and can be easily explained by
  demagnetizing currents circulating the outer tube.
\end{abstract}

\pacs{61.46.+w,76.60.Cq,71.15.Mb}

\maketitle

\section{Introduction}

Carbon nanotubes are indubitably among of the most interesting
materials discovered in the past
years\cite{iijima:1991,dresselhaus}. From a theoretical point of view,
their quasi one-dimensional character imposes quite exotic electronic
properties. In fact, some time ago it was proposed that carbon
nanotubes might be good candidates for the elusive Luttinger
liquids\cite{kane:1997}.  From a more practical point of view, carbon
nanotubes turned out to be extraordinary materials, as they possess
remarkable mechanical and electronic properties. This makes them
probable building blocks for future high-stiffness materials and
molecular nanodevices\cite{baughman:2002}. The electronic properties
of single-wall carbon nanotubes (SWCNT) can be tuned by varying the
two integers $(n,m)$ which determine the chirality of the tube. For
instance nanotubes are either metallic (for $n=m$) or semiconducting.
In the latter case, the band gap can range from more than 1\,eV [e.g.,
1.31\,eV for the (4,3) tube\cite{zolyomi:2004}] to nearly zero [e.g.,
for tubes with $\ell =0$, where $\ell= \mod(n-m, 3)$ labels the
``family'' to which the SWCNT belongs].

%To characterize experimental samples of pristine nanotubes, scientists
%employ a wide range of techniques. Electronic microscopies can yield information
%on individual tubes, such as diameter and length. For bulk samples, Raman
%spectroscopy is probably the most widely used tool.  In particular, a good
%estimate of the average diameter of the tubes can be obtained through Raman
%spectroscopy by measuring the frequency of the radial breathing
%mode\cite{kuzmany:2001}. However, it is difficult to discern the local atomic
%structure from Raman spectroscopy. This problem is particularly relevant for
%chemically modified (functionalized) nanotubes. Functionalizing nanotubes has
%attracted considerable interest, as it can facilitate their manipulation,
%enhance their solubility\cite{chen:1998}, and make them more amenable to composite
%formation.  However, to unveil the microscopic structure of these interesting
%systems (i.e., where and how the functional groups attach to the nanotube) one
%usually resorts to total energy numerical calculations\cite{lee:2005}.

Among various experimental techniques that can be used to study and
characterize materials, nuclear magnetic resonance (NMR) stands out as
being particularly suited to study local electronic
structures\cite{grant}. NMR is widely used in structural chemistry and
in material science.  It measures the local magnetic fields on the
nuclei generated by the response of the electrons to an external
uniform magnetic field. In nonmagnetic insulators this response (the
chemical shift) is determined by the orbital motion of the electrons,
and provides information about the local atomic environment
(coordination and geometry) around each nucleus.

In spite of the usefullness of the technique, only a few groups have up to now
used NMR to study carbon nanotubes. The reason is that these experiments are
quite difficult to perform, due to the presence of magnetic impurities (like Rh,
Pt, Ni, etc.) in the samples. These impurities, which are added as a catalyst
during the production of the nanotubes, lead to very broad magic-spinning angle
(MAS) spectra, with widths of about
30--50\,ppm\cite{goze-bac:2001,goze-bac:2002,hayashi:2003}. However, we expect
this situation to change soon, due to new production techniques that use
water\cite{hata:2004} or alcohol\cite{inoue:2005} to enhance the activity and
the lifetime of the catalysts. In this way, it is possible to produce samples
with a SWCNT/catalyst weight ratio 100 times higher than previous
techniques\cite{hata:2004}, which will clearly allow the measurement of NMR
spectra to unprecedented accuracy. Another possible approach involves the
removal of the ferromagnetic impurities in solution. This very promising path
has been used recently to study the NMR response of functionalized (soluble)
nanotubes\cite{alex:2005}.

In this paper, we perform a {\it ab initio} study of the magnetic response
properties of {\it infinite} carbon nanotubes. In particular, we provide
benchmark calculations of the magnetic susceptibility and NMR $^{13}$C chemical
shift of semiconducting SWCNT, both isolated and in bundles. Furthermore, we
present results for double-wall tubes.

\section{Computational details}

We use the recently proposed gauge-including projector augmented-wave (GIPAW)
approach for the calculation of NMR response in periodic systems\cite{NMRmauri}.
This approach is based on a linear response formulation of density functional
theory (DFT), as implemented in the plane-wave pseudo-potential code
PARATEC\cite{paratec}.  Contributions to the NMR spectra from the core regions
are taken into account using the projector augmented-wave (PAW) electronic
structure method\cite{blochl:1993}. We would like to remark that the current
implementation of the GIPAW method only allows the study of semiconducting
systems. The extension to metals is currently on the way, and first results are
expected shortly.

We restrict our study to $(n,0)$ (zigzag) nanotubes, belonging to the two
families characterized by $\ell = 1$ ($n=10$, 13, 16, 19) and $\ell=2$ ($n=8$,
11, 14, 17, 20). For each of these families, the band gap decreases as $1/D$,
where $D$ is the diameter of the tube. (Note that zigzag tubes have a relatively
small number of atoms in the unit cell, which eases the computational burden.)
In all calculations, wave-functions were expanded in plane-waves with a cutoff
of 40\,Ry, and we used the PBE generalized gradient approximation to the
exchange and correlation functional of DFT. For carbon, we used a norm
conserving Troullier-Martins\cite{troullier:1992} pseudo-potential with a core
radius of 1.75\,a.u. The GIPAW reconstruction included two projectors per
angular momentum for both $s$ and $p$ electrons.

The ``isolated'' tubes were rolled from a graphene sheet with ${\rm C}-{\rm
  C}=1.43$\,\AA\ and were set in a tetragonal super-lattice, with a minimum
separation of 5.5\,\AA\ between neighboring tubes. For the smaller tubes a
sampling of 20 k-points in the periodic dimension proved sufficient to converge
our results to better than 0.25\,ppm, while larger tubes required 40 k-points at
most. We modeled bundles by arranging nanotubes on a hexagonal lattice, with an
inter-tube spacing of 3.3\,\AA. In the plane perpendicular to the tube axis, we
used a $4\times4$ sampling of the irreducible Brillouin zone. Overall, we
believe that our numerical procedure yields magnetic susceptibilities to a
couple of percent and isotropic shifts with a precision better than 1\,ppm.

When calculating the chemical shift tensor, special care has to be taken with
its macroscopic component $\sigma({\bm G} = 0)$.  This is not a bulk quantity,
and its value depends on the shape of the macroscopic sample. For example, for a
spherical or cylindrical sample, and for a susceptibility tensor diagonal in the
Cartesian axis, this quantity can be related to the volumetric macroscopic
susceptibility $\chi_V$ through the relation (in cgs units)
\begin{equation}
  \sigma({\bm 0}) = -4\pi \alpha \chi_V
  \,,
\end{equation}
where $\alpha$ is a diagonal matrix with values $\alpha_{ii}=2/3$ for a sphere
and $\alpha_{xx}=\alpha_{yy}=1/2$ and $\alpha_{zz}=1$ for a cylinder (in this
latter case $z$ denotes the direction of the axis of the cylinder). In the case
of a bundle of nanotubes, we believe that the most correct choice is the
cylinder. For the isolated tubes the volumetric macroscopic susceptibility
decays as $1/A^2$, where $A$ is the distance between the centers of two adjacent
tubes. We are therefore free to choose the matrix $\alpha$ in order to speed up
the convergence of the chemical shift with $A$. We found empirically that the
choice $\alpha_{xx}=\alpha_{yy}=1/2$ and $\alpha_{zz}=2/3$ provided a faster
convergence rate than either the sphere or the cylinder.

Experimental results for the chemical shift are usually given relatively to the
isotropic shift of tetramethylsilane (TMS).  The reference TMS shielding tensor
$\sigma_{\rm TMS}$ is usually calculated within the same numerical framework
used for the systems under study. This is, e.g., the method used in
Ref.~\onlinecite{zurek:2004}. In this work, we follow a slightly different
approach. Within our GIPAW method we calculate the $^{13}$C shielding tensor of
benzene, and use the formula
\begin{equation}
  \delta^{\rm TMS}_{\rm tube} = -(\sigma_{\rm tube}
  - \sigma^{\rm iso}_{\rm benzene})
  + \delta^{\rm TMS}_{\rm benzene}
  \,,
\end{equation}
where $\sigma_{\rm tube}$ is the {\it calculated} shielding tensor of the tube
and $\sigma^{\rm iso}_{\rm benzene}$ is the average value of the diagonal of the
{\it calculated} shielding tensor of benzene. Finally, $\delta^{\rm TMS}_{\rm
  benzene}$ is the {\em experimental} isotropic chemical shift in the gas-phase
of benzene relative to TMS. As the magnetic response of the tubes is very
similar to the response of benzene, we expect in this way to minimize the
systematic errors in our calculations. The experimental $\delta^{\rm TMS}_{\rm
  benzene}$ was taken to be 126.9\,ppm\cite{jameson:1987}, while $\sigma^{\rm
  iso}_{\rm benzene}$ calculated at the equilibrium geometry was 40.0\,ppm.

\section{Magnetic Susceptibility}

\begin{figure}[t]
  \begin{center}
    \includegraphics[width=8cm,clip]{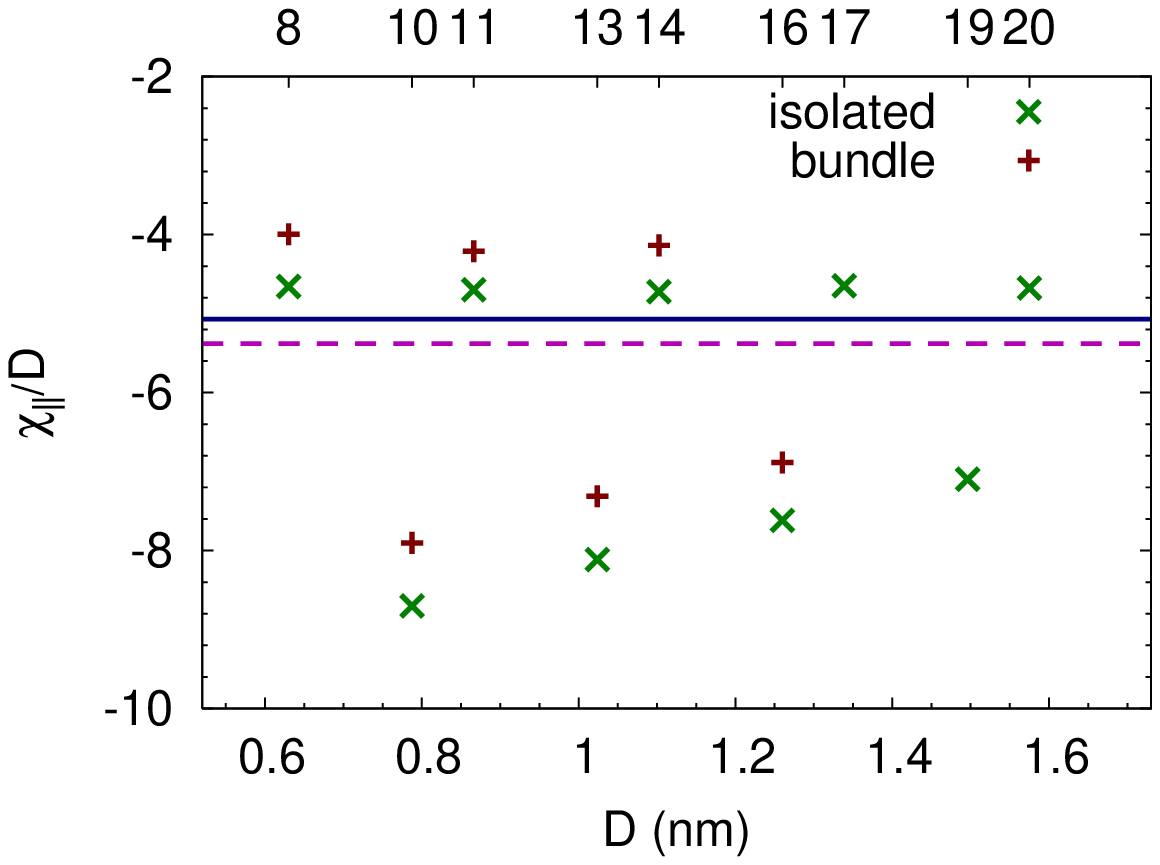}

    \vspace{0.5cm}
    \includegraphics[width=8cm,clip]{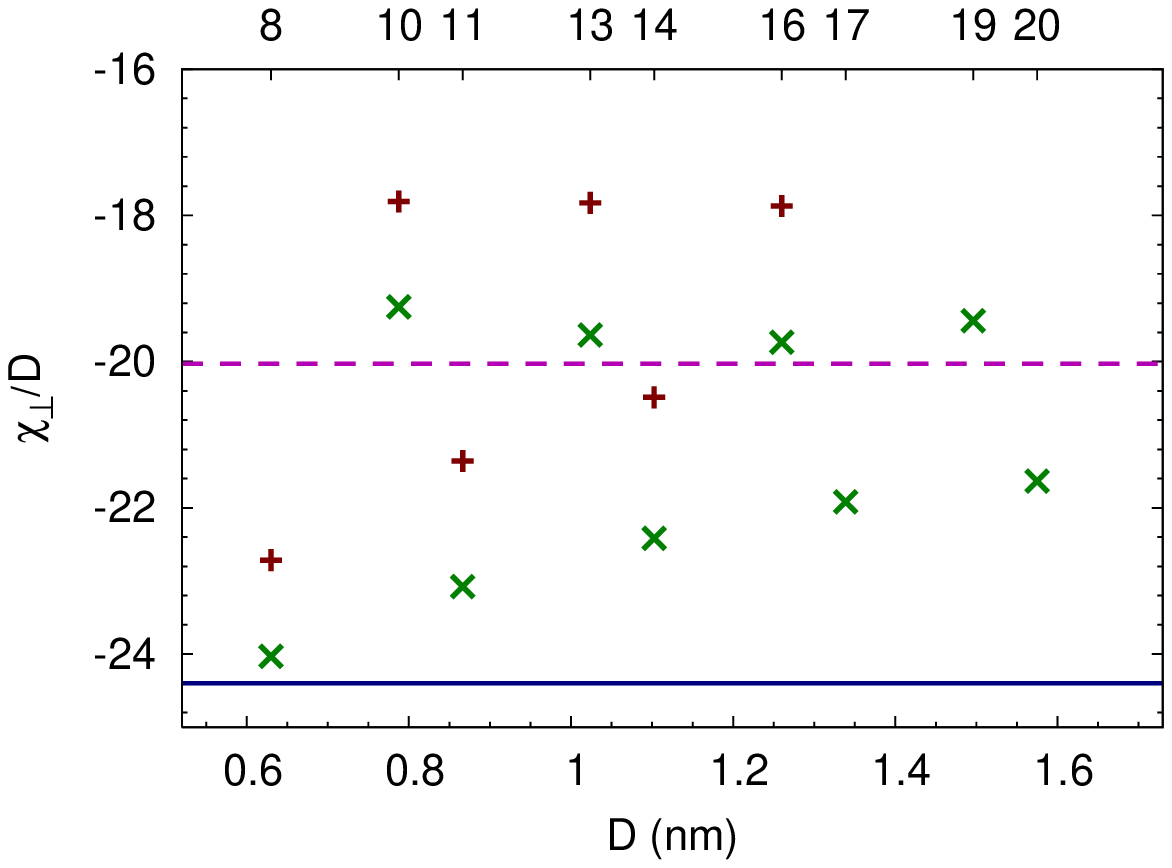}
  \end{center}
  \caption{\label{Fig:suscep1}
    (Color online) Magnetic susceptibility of zigzag semiconducting nanotubes,
    both isolated and in bundles, {\it vs} diameter ($D$) of the tube. The blue
    (solid) lines correspond to the results of
    Ref.~\protect\onlinecite{ajiki:1995} ($\chi_\parallel/D=-5.1$ and
    $\chi_\perp/D = -24.4$) and the violet (dashed) lines correspond to
    Ref.~\protect\onlinecite{lu:1995} ($\chi_\parallel/D=-5.4$ and $\chi_\perp/D
    = -20.0$). The upper axis labels the chirality $(n,0)$ of the tube. All
    results are in 10$^{-6}$cgs/(nm\,mol).}
\end{figure}

The magnetic susceptibility ($\chi$) of SWCNTs has already been the subject of
two theoretical studies: (i)~Ajiki and Ando\cite{ajiki:1995} using the ${\bm k}.{\bm
  p}$ perturbation method, and (ii) Lu\cite{lu:1995} using a tigh-binding approach
within the London approximation. It was found that $\chi$ increases linearly
with the nanotube diameter. Furthermore, for an applied magnetic field parallel
to the tube axis, the response $\chi_\parallel$ was paramagnetic for tubes with
$\ell=0$, while for tubes with $\ell=1$ or 2 the response was diamagnetic.  In
contrast, for a field perpendicular to the axis of the tube, all nanotubes were
diamagnetic.  Finally, it was found that the response of the tubes with $\ell=1$
and $\ell=2$ was almost identical.

The experimental situation is somehow more complicated, as the experimental
samples normally contain a mixture of tubes of different chiralities, including
both metallic and semiconducting tubes.  However, a recent
experiment\cite{zaric:2004} estimated the magnitude of the magnetic
susceptibility anisotropy $\chi_\parallel-\chi_\perp$ to be $\sim 14 \times
10^{-6}$\,cgs/mol for 1\,nm diameter semiconducting tubes, similar to the
predicted values of $19 \times 10^{-6}$\,cgs/mol (Ref.~\onlinecite{ajiki:1995})
and $15 \times 10^{-6}$\,cgs/mol (Ref.~\onlinecite{lu:1995}).

In Fig.~\ref{Fig:suscep1} we plot our {\it ab initio} results for the magnetic
susceptibility tensor of the isolated tubes. In contrast with
Refs.~\onlinecite{ajiki:1995} and~\onlinecite{lu:1995} we can clearly
distinguish the two families of semiconducting zigzag tubes. For the family with
$\ell=2$, $\chi_\parallel/D$ is very weakly dependent on $n$, while for the
other family ($\ell=1$) the absolute value of $\chi_\parallel/D$ decreases
slowly with the diameter of the tube. For $\chi_\perp$ the situation is
reversed: This quantity is almost independent of $n$ for $\ell=1$, and its
absolute value decreases monotonically for $\ell=2$. Note that, if we extrapolate
our results to $D\rightarrow\infty$, we find values quite consistent with
Ref.~\onlinecite{lu:1995}.

For a 1\,nm diameter tube, we obtain for the magnetic susceptibility anisotropy
($\chi_\parallel-\chi_\perp$) values between $\sim 12$ and $\sim 18 \times
10^{-6}$\,cgs/mol depending on the chirality of the tube. This result is in
quite good agreement with the value $\chi_\parallel-\chi_\perp=14 \times
10^{-6}$\,cgs/mol measured in Ref.~\onlinecite{zaric:2004}.

Regarding the bundles, we can see from Fig.~\ref{Fig:suscep1} that their
magnetic susceptibility is very similar to the one of isolated tubes. The main
difference is a (fairly rigid) decrease of $\chi_\parallel/D$ by around $0.5
\times 10^{-6}$\,cgs/(nm\,mol) and of $\chi_\perp/D$ by around $1 \times
10^{-6}$\,cgs/(nm\,mol).

\section{Chemical Shift}

\begin{figure}[t]
\begin{center}
\includegraphics[width=8cm,bb=50 50 410 302,clip]{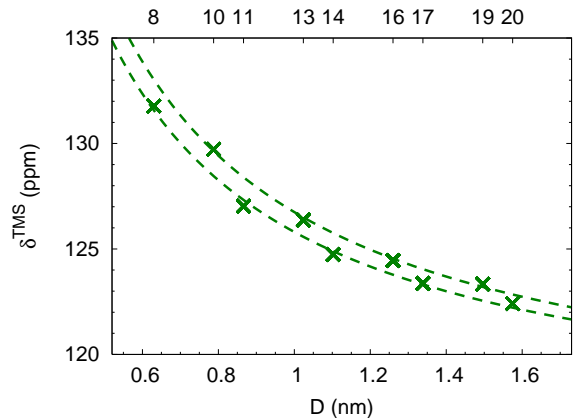}
\caption{\label{Fig:spectra1}
  (Color online) Isotropic chemical shifts for the isolated zigzag nanotubes,
  versus diameter ($D$) of the tube. The lines are the fits to the numerical
  results given by Eq.~\eqref{eq:fit}.}
\end{center}
\end{figure}

\begin{figure*}[t]
\begin{center}
\epsfig{width=13cm,clip,figure=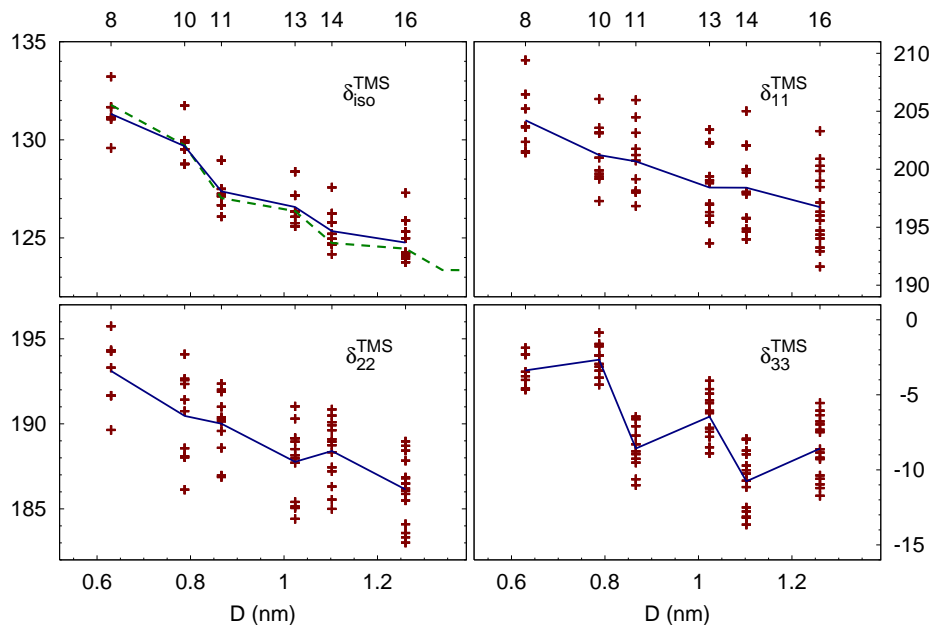}
\caption{\label{Fig:spectra2}
  (Color online) Chemical shifts relative to TMS for the zigzag nanotubes in
  bundles versus radius of the tube. The blue (solid) lines connect the average
  values for each tube. The green (dashed) line is the results for the
  isolated tubes presented before.}
\end{center}
\end{figure*}

Only a few groups have studied carbon nanotubes {\it via} NMR, whether
experimentally or theoretically. Experimental $^{13}$C MAS spectra of pristine
tubes yield isotropic shifts\footnote{The isotropic chemical shift is defined as
  the trace of the chemical shift tensor divided by 3, i.e., $\siso =
  (\saa+\sbb+\scc)/3$. The individual components are usually assigned such that
  $|\scc-\siso|\ge|\saa-\siso|\ge|\sbb-\siso|$. The magnitude of the chemical
  shift anisotropy is $\Delta\sani = (3/2)(\scc-\siso)$ and the asymmetry factor
  $\eta_{\rm C}=(\sbb-\saa)/(\scc-\siso)$.} $\siso$ of
126\,ppm\cite{goze-bac:2002,goze-bac:2001}, 124\,ppm\cite{tang:2000}, and
116\,ppm\cite{hayashi:2003}. The full width at half maximum (FWHM) of the
isotropic peak is fairly large, being around
50\,ppm\cite{goze-bac:2002,goze-bac:2001} or 30\,ppm\cite{hayashi:2003}. This
width was interpreted as a distribution of isotropic lines, resulting in part
from the presence of both semiconducting and metallic tubes in the sample. The
static spectra are quite broad, spawning more than 200\,ppm. It was also noted
that the presence of residual Ni and Co magnetic particles can contribute to the further
broadening of the spectra.

The principal values of the chemical shift tensor can be obtained from the NMR
spectra through a fitting procedure. In this way, Ref.~\onlinecite{tang:2000}
found $\delta^{\rm TMS}_{11}=195$\,ppm, $\delta^{\rm TMS}_{22}=160$\,ppm, and
$\delta^{\rm TMS}_{33}=17$\,ppm (which leads to $\Delta\delta_{\rm anis} =
-160.5$\,ppm and $\eta_{\rm C}=0.33$), while in Ref.~\onlinecite{goze-bac:2002}
it was obtained $\delta^{\rm TMS}_{11}=240$\,ppm, $\delta^{\rm
  TMS}_{22}=171$\,ppm, and $\delta^{\rm TMS}_{33}=-36$\,ppm ($\Delta\delta_{\rm
  anis} = -241.5$\,ppm and $\eta_{\rm C}=0.43$). The large difference between
the two experiments (see Fig.~\ref{Fig:static}) can be understood from the
difficulty of obtaining good samples.

On the theoretical side, the first attempt at describing the NMR response of
carbon nanotubes\cite{latil:2001} was performed within a tight-binding scheme
and using the London approximation (i.e., retaining only the London ring-current
contribution to the chemical shift and neglecting the Pople
correction\cite{pople:1962}). To justify this approximation it was argued
that the Pople contribution gave rise to a global shift, independent of
chirality, that could not be used to distinguish between different tubes. Within
this framework, it was estimated that the difference in the isotopic chemical
shift between metallic and semiconducting tubes amounted to 12\,ppm. (Absolute
shifts were not given due to the neglection of the Pople correction.)
Furthermore, it was found that the NMR of both metallic and semiconducting is
mostly insensitive to diameter and chirality.

More recently the $^{13}$C chemical shift of finite-size (9,0) single wall
carbon nanotubes was investigated\cite{zurek:2004}. The tube fragments studied
were capped by either C$_{30}$ or hydrogen. For the longest fragment it was found
$\delta_{\rm iso}^{\rm TMS} \approx 133$\,ppm for carbon atoms at the center of
tube, in reasonably good agreement with experimental data. For the
C$_{30}$-capped tubes, apparently the best models for the infinite systems, the
shifts in the middle of the fragment were invariably smaller than those at the
end (where they reached $\delta_{\rm iso}^{\rm TMS} \approx 155$\,ppm).

\subsection{Isotropic shift}

Our results for the isolated tubes are presented in Fig.~\ref{Fig:spectra1}.
The two families can be clearly identified: The family with $\ell=1$ has a
slightly larger chemical shift than the family with $\ell=2$. The results can be
fitted by the relation
\begin{equation}
  \label{eq:fit}
  \delta_{\rm iso}^{\rm TMS} = \left\{\begin{array}{ccc} 
    10.70/D + 116.0 & {\rm if}  & \ell = 1 \\ 
    \phantom{0}9.83/D  + 116.0 & {\rm if}  & \ell = 2
  \end{array}\right.
  \,,
\end{equation} 
where $D$ is the diameter of the nanotube in nanometers, and $\delta_{\rm
  iso}^{\rm TMS}$ is the isotropic chemical shift relative to TMS in ppm.
According to this formula, in the limit of infinite radius the isotropic
chemical shift tends to 116.0\,ppm. This value compares reasonably well with the
estimate of 128\,ppm for $\delta^{\rm TMS}_{\rm graphene}$ of a perfect graphene
layer (with zero density of states at the Fermi
level)\cite{lauginie:1988,goze-bac:2002}.  For a nanotube with a diameter of
1.2\,nm, like the ones common in experimental samples, we obtain a
$\delta_{\rm iso}\approx 125$\,ppm, in excellent agreement with
experiment\cite{goze-bac:2002,goze-bac:2001,tang:2000,hayashi:2003}. Note that,
as already indicated in Refs.~\onlinecite{goze-bac:2001}
and~\onlinecite{goze-bac:2002}, the NMR chemical shift depends very weakly on
the diameter and on the chirality of the semiconducting nanotubes: the isotropic
shifts of the zigzag tubes studied in this work lie between 122\,ppm for the
smallest tube [the $(8,0)$] to 132\,ppm for the $(20,0)$ tube.

The isotropic shifts of the bundles are represented by red crosses in
Fig.~\ref{Fig:spectra2}. Since the hexagonal lattice breaks the chiral symmetry
of the tubes, the carbon atoms around the circumference of the tube are no
longer equivalent, resulting in a slight spread of $\delta^{\rm TMS}$ in the
unit cell. For the nanotubes studied here, this spread of the isotropic shift
was quite small (around 3\,ppm), and fairly independent of the radius of the
tube. It can also be seen that the average shift in a bundle is nearly equal to
the isotropic value of the isolated tube\cite{goze-bac:2002}.

We are now in condition to discuss the large width MAS spectra, that amounts to
30--50\,ppm. Several effects can contribute to this width:

(i)~Experimental samples are composed of nanotubes of different diameters and
chiralities. Assuming that the spread of diameters is around 0.5\,nm, and the
distribution is centered around 1.2\,nm, from Fig~\ref{Fig:spectra1} we obtain a
spread of around 4\,ppm. Furthermore, the difference in chemical shift between
semiconducting and metallic tubes was estimate to be 12\,ppm\cite{goze-bac:2002}.

(ii)~As experiments are usually performed with samples of SWCNTs arranged in
bundles, we have to take into account the spread of isotropic chemical shifts
due to the breaking of chiral symmetry.  As we saw from Fig.~\ref{Fig:spectra2}
this amounts to around 3\,ppm for zigzag tubes.

(iii)~As mentioned before, in order to obtain the macroscopic component of the
chemical shift tensor, we assumed that the bundle was of cylindrical shape.
However, in a real sample, the bundle can assume shapes quite different from a
perfect cylinder. To address this issue, we performed calculations assuming a
spherical shape for the bundle. For the (14,0) tube, a typical nanotube of
1.1\,nm diameter, the isotropic shift calculated for the cylinder was
125.3\,ppm, while the sphere gave 122.4\,ppm.  Note that, even if the
difference, 3\,ppm, is much smaller than the experimental width of the MAS
spectrum, it has nevertheless the same magnitude as the difference between
nanotubes of similar diameter.

(iv)~Finite-size SWCNTs can be either open, or capped by fullerene-like
structures. It was found in Ref.~\onlinecite{zurek:2004} that the carbon atoms
close to the ends of these (finite) tubes can have an isotropic shift up to
20\,ppm larger than the atoms far from the ends. However, although not infinite,
the SWCNTs present in experiment are quite long, so we do not expect this effect
to contribute significantly to the NMR spectra.

(v)~Other effects: Clearly every experimental sample of SWCNTs contains defects
and magnetic impurities which will contribute to the broadening of the MAS peak.
Furthermore, there may be interactions of the semiconducting nanotubes with the
conduction electrons of metallic tubes contained in the bundle, etc.

We can see that effects (i)--(iii) can contribute around $\sim 20$\,ppm to the
width of the MAS peak. We therefore expect that the remaining $\sim$10--30\,ppm
of the experimental width\cite{goze-bac:2001,goze-bac:2002,hayashi:2003} are due
to (v). This means that by using better samples with a much lower concentration
of magnetic impurities\cite{hata:2004,inoue:2005} the resolution of $^{13}$C NMR
spectra can be improved by at most a factor of 2.

\subsection{Anisotropic shifts}

\begin{figure}[t]
\begin{center}
\epsfig{width=7cm,clip,figure=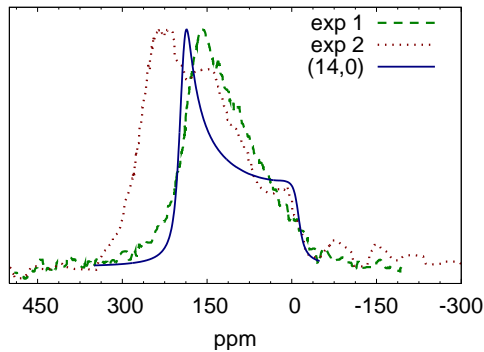}
\caption{\label{Fig:static}
  (Color online) Experimental versus simulated static spectra for the
  carbon nanotubes. The curve labeled 'exp 1' was taken from
  Ref.~\protect\onlinecite{tang:2000}, while 'exp 2' was taken from
  Ref.~\protect\onlinecite{goze-bac:2002}. The theoretical curve
  corresponds to the static spectrum of a bundle of $(14,0)$
  nanotubes, convoluted with a Lorentzian of 30\,ppm FWHM.}
\end{center}
\end{figure}

Also in Fig.~\ref{Fig:spectra2} we show the principal values of the chemical
shift tensor for the bundles. The first principal value, $\delta^{\rm TMS}_{11}$
corresponds to the longitudinal (along the tube axis) direction, $\delta^{\rm
  TMS}_{22}$ lies on the orthoradial direction, and $\delta^{\rm TMS}_{33}$ is
the radial component of the chemical shift tensor. As expected, the chemical
shift is extremely anisotropic, with the longitudinal and orthoradial directions
corresponding to the in-plane directions of a graphene sheet (or of a benzene
molecule), and the radial component corresponding to the direction perpendicular
to the sheet. As for the isotropic shift, the individual principal values are
spread due to the interaction between the tubes in the bundle, but this time by
as much as 10\,ppm. Note that Ref.~\onlinecite{goze-bac:2002} found a value of
20\,ppm for the broadening of the tensor due to the influence of the first
neighbor tube (and quasi-independent on the electronic properties of this first
neighbor). There is, therefore, a partial cancellation when we include all the
six neighbors in the hexagonal lattice. Note that this 10\,ppm spread is much
larger than the 3\,ppm spread of the isotropic line. It is also much larger than
the difference between the two families of nanotubes we have studied.

In Fig.~\ref{Fig:static} we compare a simulation of the static spectrum of a
$(14,0)$ tube with experimental results. The theoretical spectrum shows a
typical powder line-shape with principal values $\delta^{\rm
TMS}_{11}=198$\,ppm, $\delta^{\rm TMS}_{22}=188$\,ppm, and $\delta^{\rm
TMS}_{33}=-11$\,ppm ($\Delta\delta_{\rm anis} = -204$\,ppm and $\eta_{\rm
C}=0.07$). The two experimental curves are quite different from each other, so
care must be taken when comparing theoretical to experimental spectra. In
general, we believe that our results are in quite good agreement with
experiments. The main difference turns out to be the value for the asymmetry
factor, which is considerably smaller in our calculations than in the
experimental works.

\section{Double-Wall Tubes}

Recently, Simon {\it et al.} studied double-wall nanotubes using
NMR~\cite{simon:2005}. Their samples were produced through the high-temperature
annealing of isotope enriched fullerenes encapsulated in single-wall tubes. In
this way, they were able to measure directly the chemical shift of the inner
tube. They found an isotropic shift of 111\,ppm, which is significantly smaller
than the isotropic shift of larger single wall tubes. The authors pointed out
two possible reasons for this difference: (i)~the stronger curvature of the
inner tube, or (ii)~diamagnetic shielding due to the outer tube.
Note that in our calculations the chemical shift {\it increases} with decreasing
radius (see Fig.~\ref{Fig:spectra1}), which rules out (i).

To study this problem, we performed {\it ab initio} calculations of the isolated
double-wall tube (8,0)@(16,0). We obtained for the isotropic chemical shift of
the atoms composing the inner tube a value of 111\,ppm [to be compared with
132\,ppm for the isolated (8,0) tube], implying a diamagnetic shift due to the
outer tube of 21\,ppm. On the other hand, the outer tube had an isotropic
chemical shift of 126\,ppm [to be compared to 124\,ppm of the isolated (16,0)
tube]. Note that our {\it ab initio} result for the inner tube (111\,ppm)
coincides with the value measured in Ref.~\onlinecite{simon:2005}.

In order to understand our results, and to study the effect of diamagnetic
shielding, we will resort to a simple classical model. Let us imagine that the
outer nanotube has the form of a long hollow cylinder with axis parallel to
$\hat z$. We will distinguish two cases, when the probing magnetic field ${\bm
  B}_0 \parallel \hat z$ and when ${\bm B}_0 \perp \hat z$.  In the former case,
the induced current $\bm j$ on the surface of the cylinder can be written (in
linear response) as ${\bm j}=\beta_\parallel B_0 \hat \theta$, where
$\beta_\parallel$ is a constant. The magnetization per unit length can then be
calculated from (in cgs units)
\begin{equation}
  {\bm M} = \frac{1}{2 c L} \int\!\! d^3r\; {\bm r}\times {\bm j}({\bm r})
  = \frac{\beta_\parallel B_0}{c} \frac{\pi D^2}{4} \hat{z}
  \,,
\end{equation}
where $c$ denotes the speed of light, $L$ is the lattice parameter in the
periodic direction, $D$ is the diameter of the cylinder, and $\hat{z}$ is the
unit vector of direction $z$. The volume of integration is one unit cell.
Remembering that $\bm M = \frac{\pi D}{N_{\rm A}} n_{\rm C} \chi_{\parallel} \bm
B_0$, where $\chi_{\parallel}$ is the molar susceptibility parallel to the axis
of the cylinder, $N_{\rm A}$ is the Avogadro number, and $n_{\rm C}$ is the
density of carbon atoms per unit area of the nanotube ($n_{\rm
  C}=37.71$\,nm$^{-2}$), we can obtain the induced current
\begin{equation}
  \beta_\parallel = c \frac{4 n_{\rm C}}{N_{\rm A} D} \chi_{\parallel}
  \,.
\end{equation}
On the other hand, it is a simple excercise of magnetostatics to calculate the
magnetic field generated by $\bm B_0$ and by the induced current. The total
magnetic field ($\bm B_{\rm in}$) inside the cylinder (i.e., the field felt by
the inner nanotube) is constant. Furthermore
\begin{equation}
  \label{bshift}
  \frac{B_{\rm in} - B_0}{B_0} = 4\pi \frac{4 n_{\rm C}}{N_{\rm A} D}
  \chi_{\parallel}
  \,.
\end{equation}
Note that as $\chi_{V \parallel}$ is negative for our nanotubes,
Eq.~\eqref{bshift} gives rise to a {\it diamagnetic shift}. The same derivation
can be done for an external field perpendicular to the tube axis. In this case,
the current is, for a magnetization in the $\hat y$ direction, $\bm j =
\cos(\theta) \hat{z}$ on the sides of the cylinder, and $\bm j = \pm \beta_\perp
B_0 \hat{x}$ on its bases. The total (external plus induced) field inside the
tube is still constant, and it reads
\begin{equation}
  \label{bshift2}
  \frac{B_{\rm in} - B_0}{B_0} =  4\pi \frac{2 n_{\rm C}}{N_{\rm A} D}
  \chi_{\perp}
  \,.
\end{equation}
We can now calculate the diamagnetic shift in the inner tube due to the presence
of the outer tube. As a typical example we chose the double-wall tube
(8,0)@(16,0). Inserting its diameter (1.26\,nm) and the calculated
susceptibility ($\chi_\parallel = -9.60 \times 10^{-6}$\,cgs/mol and
$\chi_\perp = -24.86 \times 10^{-6}$\,cgs/mol) of the (16,0) tube in
Eqs.~\eqref{bshift} and~\eqref{bshift2} we obtain that the shift due to the
shielding is $\delta^{\rm shield}_{33}=-24$\,ppm and $\delta^{\rm
  shield}_{11}=\delta^{\rm shield}_{22}=-31$\,ppm. This leads to an isotropic
shift of $\delta_{\rm iso}^{\rm shield}=-29$\,ppm. From our calculations, the
isotropic shift of the isolated (8,0) tube is 131\,ppm, which leads to a total
chemical shift for the inner tube of $\sim102$\,ppm.  This is a fairly good
agreement with our {\it ab initio} results for such a crude model, and fully
justifies its use in these systems.

\section{Conclusions}

We performed calculations of the magnetic susceptibility and $^{13}$C chemical
shift for semiconducting zigzag carbon nanotubes, both isolated and in bundles.
Our results are in very good agreement with experiment. For all nanotubes
studied, the isotropic shift lies between 120\,ppm and 135\,ppm. This interval
of 15\,ppm is quite narrow, and is at the limit of the precision that NMR
experiments have obtained for these systems. However, we expect that, with the
advent of better samples, NMR will become a key tool to study and characterize
carbon nanotubes.

\section*{Acknowledgments}

The authors would like to thank S. Botti, C. Goze-Bac, and A. Rubio for fruitful
discussions.  MALM was supported by a Marie Curie Action of the European
Commission, Contract No. MEIF-CT-2004-010384. Calculations were performed at the
IDRIS supercomputing center (grant \#051202).

%\bibliography{biblio}

%  \bibitem{separate} ??? not cited ?
%  R. Krupke, F. Hennrich, H. v. Löhneysen, and M.\.M. Kappes
%  Science {\bf 301}, 344-347 (2003);
%  M. Zheng, A. Jagota, M.\,S. Strano, A.\,P. Santos, P. Barone, S.\,G. Chou, 
%  B.\,A. Diner, M.\,S. Dresselhaus, R.\,S. Mclean, G.\,B. Onoa, G.\,G. Samsonidze, 
%  E.\,D. Semke, M. Usrey, and D.\,J. Walls
%  Science {\bf 302}, 1545 (2003).

\end{document}